# Aligning van der Waals heterostructures using electron backscatter diffraction

R. Bangari[1], M. Mosayebi[2,3], J. Buchner[4], J.D. Caldwell[4], N. Bassim[2,3], and T. G. Folland [1]

[1]Department of Physics and Astronomy, The University of Iowa, Iowa City 52242, IA, USA

[2]Materials Science and Engineering Department, McMaster University, Hamilton L8S 4L8, ON, Canada

[3]Canadian Centre for Electron Microscopy, McMaster University, Hamilton, Ontario L8S 4M1, Canada

[4]Department of Mechanical Engineering, Vanderbilt University, Nashville 37235, TN, USA

## Abstract

Precise and accurate determination of crystallographic orientation is crucial for engineering van der Waals heterostructures, where the twist angle between layers controls emergent electronic and optical properties. While Electron Backscatter Diffraction (EBSD) has been extensively used for bulk materials, its application to van der Waals materials remains largely unexplored. In this work, we demonstrate EBSD as a robust and versatile tool for determining crystallographic orientations of van der Waals materials with high precision. We show quantitative agreement between EBSD-determined orientations and facet orientations in orthorhombic α-MoO$_3$ flakes on silicon substrates. We use Grain Reference Orientation Distribution (GROD) and Kernel Average Misorientation (KAM) across the flakes to demonstrate precision better than 0.2°. We extend this technique to other low-symmetry materials, specifically, monoclinic α-As$_2$Te$_3$, monoclinic GaTe and triclinic ReSe$_2$, demonstrating broad applicability across van der Waals materials with different crystal structures. Finally, as a proof-of-concept application, we leverage EBSD-determined orientations to engineer twisted α-MoO$_3$ heterostructure with precisely controlled twist angle, enabling observation of recently reported canalized phonon polaritons. Our results establish EBSD as a powerful characterization method for van der Waals materials, enabling precise orientation control essential for twistronics and twist-optics.

Van der Waals (vdW) materials have garnered great interest for their layered crystalline structure that allows them to be exfoliated down to monolayer thicknesses. Within this vast library, low-symmetry crystals like orthorhombic α-$MoO_3$ [1, 2], monoclinic α-$As_2Te_3$ [3, 4], monoclinic GaTe [5, 6] and triclinic $ReSe_2$ [7, 8] are particularly intriguing due to their highly anisotropic electronic, thermoelectric, opto-electronic and optical properties [9-13]. The absence of strict lattice matching requirements allows these diverse layers to be mechanically stacked into heterostructures, where the relative twist angle between layers serves as a critical degree of freedom [14]. This tunability has given rise to the fields of twistronics [15, 16] and twist-optics [17, 18], where the precise alignment of crystallographic axes dictates the emergence of moiré superlattices [19], correlated insulating states [20], and topological photonic transitions [21, 22].

Fully harnessing emergent phenomena in twisted vdW stacks requires exceptional control over crystallographic orientation. In electronic systems like twisted bilayer graphene, the magic angle (1.1°) demands a tolerance of less than 0.1° to preserve the flat-band condition essential for superconductivity; even minute twist-angle disorder can suppress correlated states [23]. Similarly, in low-symmetry optical systems like twisted α-$MoO_3$, the transition to canalization, where phonon polaritons (PhPs) propagate without diffraction occurs at specific twist angles with tolerances of a few degrees [24]. Current characterization methods struggle to meet these demands. Optical techniques such as angle-resolved polarized Raman spectroscopy and second harmonic generation are diffraction-limited (~500 nm) and have accuracies generally only up to 0.5° [25-29]. Furthermore, the selection rules for Raman and nonlinear processes present challenges as a general-purpose characterization technique for any material. An alternative is visual alignment to straight facet edges, which often align with the crystallographic axes. However, these can be unreliable as cleavage planes don't always align with crystallographic axes in many vdW flakes. Finally, for twisted stacking of identical materials, "tear-and-stack" and "cut-and-stack" methods are used due to the inability to determine the crystal orientation by optical characterization [30]. Even though these techniques provide high-precision twist angles, small sample size is unavoidable due to halving of the size of the exfoliated parent monolayer and does not allow stacking of different materials or flakes from different substrates [31].

To bridge this gap between fabrication requirements and characterization capabilities, Electron Backscatter Diffraction (EBSD) offers a powerful solution. EBSD is a crystallographic characterization technique that utilizes a scanning electron microscope (SEM) to analyze diffraction patterns generated by electrons backscattered from crystalline samples. When the electron beam interacts with a crystalline material, a fraction of the incident electrons undergo backscattering from the crystal lattice planes that impinge upon a phosphor screen, generating a characteristic diffraction pattern known as a Kikuchi pattern [32]. While EBSD has been traditionally used to find crystallographic makeup of metals, alloys, ceramics and minerals [33-35], and has been shown to provide high spatial resolution down to 50 nm [36], it has been explored as a tool for characterizing vdW materials only recently [37, 38].

In this work, we systematically study the accuracy and precision of EBSD measurements in determining the crystallographic orientation of orthorhombic α-MoO$_3$ flakes showing excellent agreement with facet orientation and angular precision better than 0.2°. We then show the effectiveness of the technique in lower symmetry vdW materials including monoclinic $\alpha$-As$_2$Te$_3$, monoclinic GaTe and triclinic ReSe$_2$. Such samples are particularly challenging, as their cleavage planes often do not align with their crystallographic axes. We further demonstrate the use of EBSD in practice by fabricating a twisted stack of α-MoO$_3$ flakes on Silicon (Si) substrates to achieve directional PhP propagation at 885 cm$^{-1}$.

The vdW samples were prepared by mechanically exfoliating bulk crystals obtained from 2D Semiconductors using adhesive tape onto Si substrates. The EBSD measurements were performed using a dual-beam Thermo Fisher Scientific Helios G5 Xe plasma focused ion beam (PFIB) system equipped with a Symmetry S2 CMOS EBSD detector (Oxford Instruments) and operated with AZtec acquisition software. Measurements were conducted at an accelerating voltage of 20 kV and a beam current of 13 nA. The working distance ranged from 14–22 mm to optimize pattern quality and detector geometry. EBSD orientation maps were acquired using step sizes between 20–500 nm, with an exposure time of 10 ms per pattern. The details of each measurement are provided in the Supporting Information (Table S1).

To validate the efficacy of EBSD for vdW materials, we first examined α-MoO$_3$ flakes. Figure 1(a) shows the anisotropic crystal structure of α-MoO$_3$. The α-MoO$_3$ lattice is composed of MoO$_6$ octahedra that form rigid zigzag chains through edge-sharing along the [001] direction, whereas these chains are connected to each other by weaker corner-sharing bonds along the [100] direction. This bonding anisotropy implies that the crystal preferentially grows or fractures parallel to the robust edge-sharing chains, resulting in rectangular structures where the long edge corresponds to the [001] [2, 39]. α-MoO$_3$ flakes therefore mostly exfoliate into flat rectangular shapes with well-defined edges, which provides a way to compare the orientation determined by EBSD and the facet orientation given by flake edges that is indicative of the crystallographic orientation. In most of the literature, twisted α-MoO3 stacks are fabricated using facet orientation as a reference for crystal orientation.

The SEM micrograph of an α-MoO$_3$ on Si sample, presented in Figure 1(b), reveals multiple flakes with varying in-plane orientations. By raster-scanning the electron beam across the sample, we generated a spatially resolved Inverse Pole Figure (IPF-Y) orientation map (Figure 1(c)). In this map, the color of each pixel corresponds to the crystallographic axis aligned with an in-plane direction (Y-direction). The uniform coloring within individual flakes confirms their single-crystalline nature, while the difference in color between different flakes shows their different in-plane orientations. The Electron Backscatter Diffraction Pattern (EBSP) from an α-MoO₃ flake (Figure 1(e)) differs markedly from that generated by the single-crystal Si substrate (Figure 1(d)), enabling clear phase discrimination. Figure 1(f) visualizes the unit cell orientations derived from this analysis. Crucially, the crystallographic orientations determined by EBSD show that the [001] and [100] crystallographic directions align with the physical edges of the flakes.

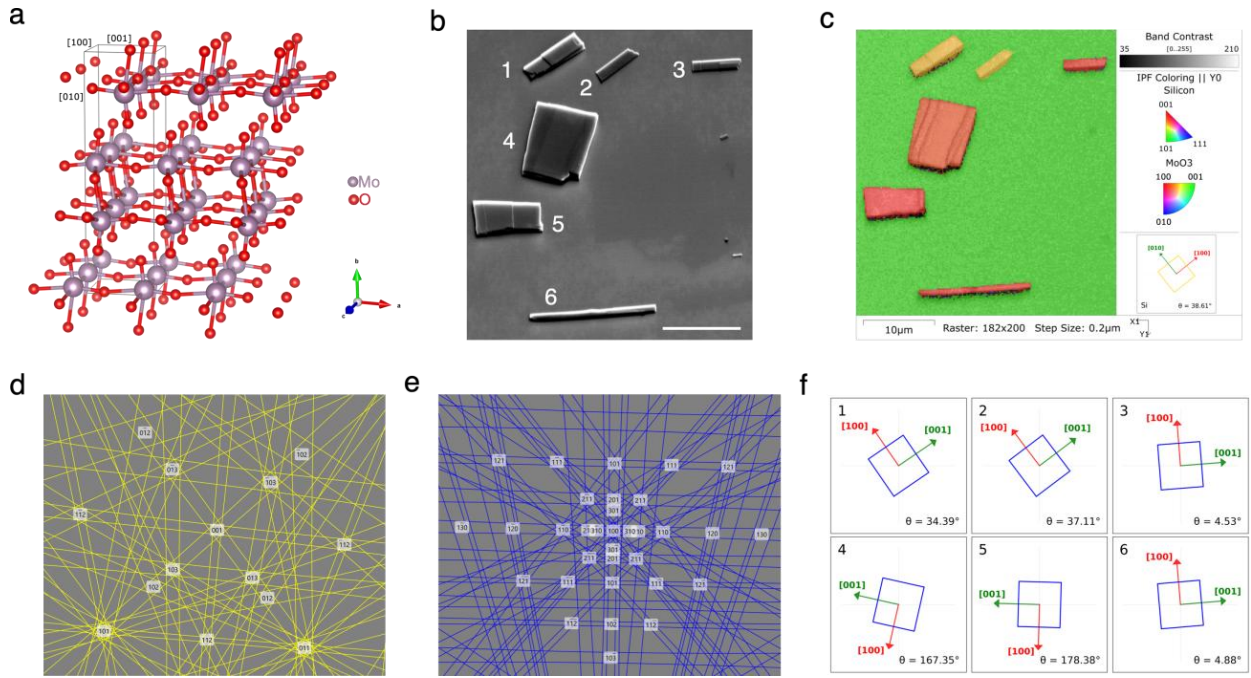

**Figure 1. EBSD orientation mapping of α-MoO₃ flakes.** (a) Crystal structure of α-MoO₃ used in the Electron Backscatter Diffraction (EBSD) characterization with the following unit cell parameters: a=3.9629 Å, b=13.8649 Å, c=3.69756 Å; α=γ=β=90°. (b) Tilted Scanning Electron Microscope (SEM) micrograph showing multiple exfoliated α-MoO₃ flakes on a Si substrate with varying in-plane orientations. The scale bar is 10 µm. (c) Inverse Pole Figure (IPF-Y) orientation map overlaid with band contrast corresponding to (b); uniform color within the flakes indicates single-crystallinity, while distinct colors represent unique in-plane orientations. The inset shows the crystal orientation of the Si substrate where angle θ represents the in-plane orientation of the [100] axis with respect to the X axis. (d–e) Representative Kikuchi diffraction patterns acquired from a single point on (d) the Si substrate and (e) an α-MoO₃ flake, displaying distinct band geometries. (f) Visualization of unit cell orientations for the indexed flakes, highlighting the alignment of the [001] and [010] axes with the physical flake edges. The angle θ depicts the in-plane orientation of the [001] axis with respect to the X axis.

To rigorously quantify the accuracy of our EBSD measurements, we compared the crystallographic orientation of α-MoO₃ flakes on Si obtained from EBSD indexed with the physical orientation of the long facets of the α-MoO₃ flakes, which are known to cleave preferentially into rectangles parallel to the [001] crystal axis. The details of the facet orientation extraction are described in Supporting Information. Figure 2(a) plots the EBSD-determined in-plane angle against the facet angle extracted from SEM images for a statistical ensemble of flakes. The data was collected on flakes of varying sizes, orientations, and thicknesses down to ~70 nm. The uncertainty in facet orientation determination ranges from 1.09° and 4.18° with an average of 2.27°. We observe a highly linear correlation between EBSD and facet angle that closely tracks the ideal 1:1 line. This confirms that the EBSD indexing provides an accurate measure of crystal orientation of the flakes consistent with their facet orientation. The errors from EBSD measurements are not included in the plot as they are significantly lower as described below.

We evaluated the angular precision of the EBSD measurements using Kernel Average Misorientation (KAM) and Grain Reference Orientation Distribution (GROD) analyses. A KAM map shows the mean of the misorientation angles $\Delta\theta_{ij}$ calculated from $\mathbf{\Delta\Omega_{ij}} = \Omega_i . \Omega_j^{-1}$, where $\Omega_i$ and $\Omega_j$ are orientations at the point being mapped and a neighboring point respectively [40, 41]. KAM values in this study are computed using a 3×3 neighborhood. On the other hand, GROD maps the misorientation angle $\Delta\theta$ between the orientation $\Omega_i$ at each map point and a reference orientation $\Omega_r$ which represents the average orientation inside the grain, providing a metric for global lattice variation within the flake [42]. GROD is generally associated with disorders in the sample, while KAM is considered to be a good metric for the resolution of the EBSD measurement as it looks at local differences in the orientation [43]. This is especially true in vdW flakes as lattice deformations would not be expected to be significant over a small neighborhood.

We obtain distributions for both the KAM and GROD metrics from EBSD measurements and extract the mean (Figure 2(b)) and median (Figure 2(c)) of the distributions over each flake. Small and only slightly varying values are observed in KAM mean and median across different samples indicating consistently high precision in our EBSD measurements. GROD values are similarly low and closely track the KAM metric, reflecting minimal crystallographic distortion within the flakes. It is known in literature that mechanically exfoliated vdW flakes show lower strain and defects compared to other synthesis methods [44]. The mean of KAM and GROD cumulative distributions from all the samples are 0.20° and 0.37° respectively, and the overall median values are 0.13° and 0.24° respectively. These values are substantially smaller than the uncertainty associated with facet orientation determination (~2.27°). We therefore find that EBSD can resolve the crystallographic orientation in our vdW samples with a precision finer than 0.2°.

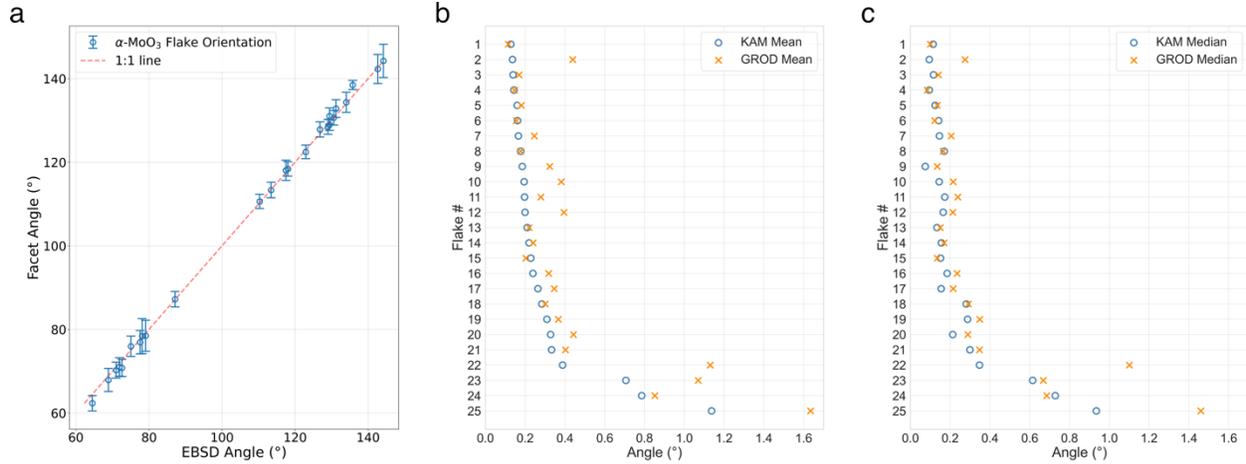

**Figure 2. Validation of EBSD accuracy and precision on α-MoO₃ flakes. (a)** Comparison of physical facet angle (extracted from SEM images) and the crystallographic in-plane angle (determined by EBSD indexing) for an ensemble of α-MoO₃ flakes on Si. The data follows a 1:1 correspondence with a weighted Pearson correlation of 0.9993 between the datasets, confirming orientation accuracy. Error bars represent the standard deviation of facet orientation determination within each flake. **(b)** Mean and **(c)** Median values of Kernel Average Misorientation (KAM) and Grain Reference Orientation Distribution (GROD) for individual flakes, with flakes sorted by ascending KAM mean. KAM represents local measurement noise, while GROD captures global orientation spread.

Having established the accuracy and precision of EBSD on orthorhombic α-MoO₃, we extend this technique to other technologically relevant vdW materials with lower crystal symmetries. We observe uniform coloring across the flakes in the EBSD orientation maps obtained from As$_2$Te$_3$ (Figure 3(a)), GaTe (Figure 3(b)) and ReSe$_2$ (Figure 3(c)) on Si substrates, indicating they are single crystalline in nature with no significant grain boundaries or twinning. In all cases, we obtained distinct Kikuchi patterns that could be reliably indexed against their respective crystallographic databases. The low mean and median values of the KAM and GROD distributions seen for the three flakes (Table 1) show high precision in the orientation characterization and minimal crystal deformation in the samples. These lower symmetry vdW materials do not necessarily have the same preferential facet alignment as the orthorhombic α-MoO$_3$ flakes and hence optical techniques such as polarized Raman are generally used to characterize the crystal axes in such materials [45, 46].

We extract the angular deviation of the principal crystal axes of the flakes from the sample surface normal (Z-axis). As shown in Figure 3(d), the [100] axis of an As$_2$Te$_3$ flake is aligned nearly parallel to the Z-axis, coinciding with the Si [111] substrate peak. The [010] and [001] axes are found approximately in the sample plane, confirming that the As$_2$Te$_3$ flake is oriented with the [100] plane parallel to the substrate surface. The orientation of the monoclinic GaTe flake is more complex (Figure 3(e)). Neither the [100], [010], nor [001] axes align directly with the surface normal. The [100] axis is offset by approximately 34.5°, and the [001] axis by approximately 40.5° from the Z-axis, while the [010] axis lies nearly in the sample plane. The triclinic ReSe$_2$ flake (Figure

3(f)) exhibits a [001] axis oriented at approximately 22.5° relative to the surface normal, with the [100] and [010] axes lying close to the sample plane.

We further visualize the approximate side views of the crystal structure of these materials adjusted using the measured tilt of the [001] crystal axes (Figure 3(g-i)). We observe that indeed the cleaving planes for all the flakes match with sample plane as vdW materials typically cleave at a plane with largest atomic distances or least interlayer forces normal to it. The out of plane orientations of the exfoliated flakes agree with those observed in literature [47-49]. This demonstration confirms that EBSD orientation determination is not limited to high-symmetry phases but is a truly universal tool capable of resolving the crystallographic identity of the most complex vdW systems.

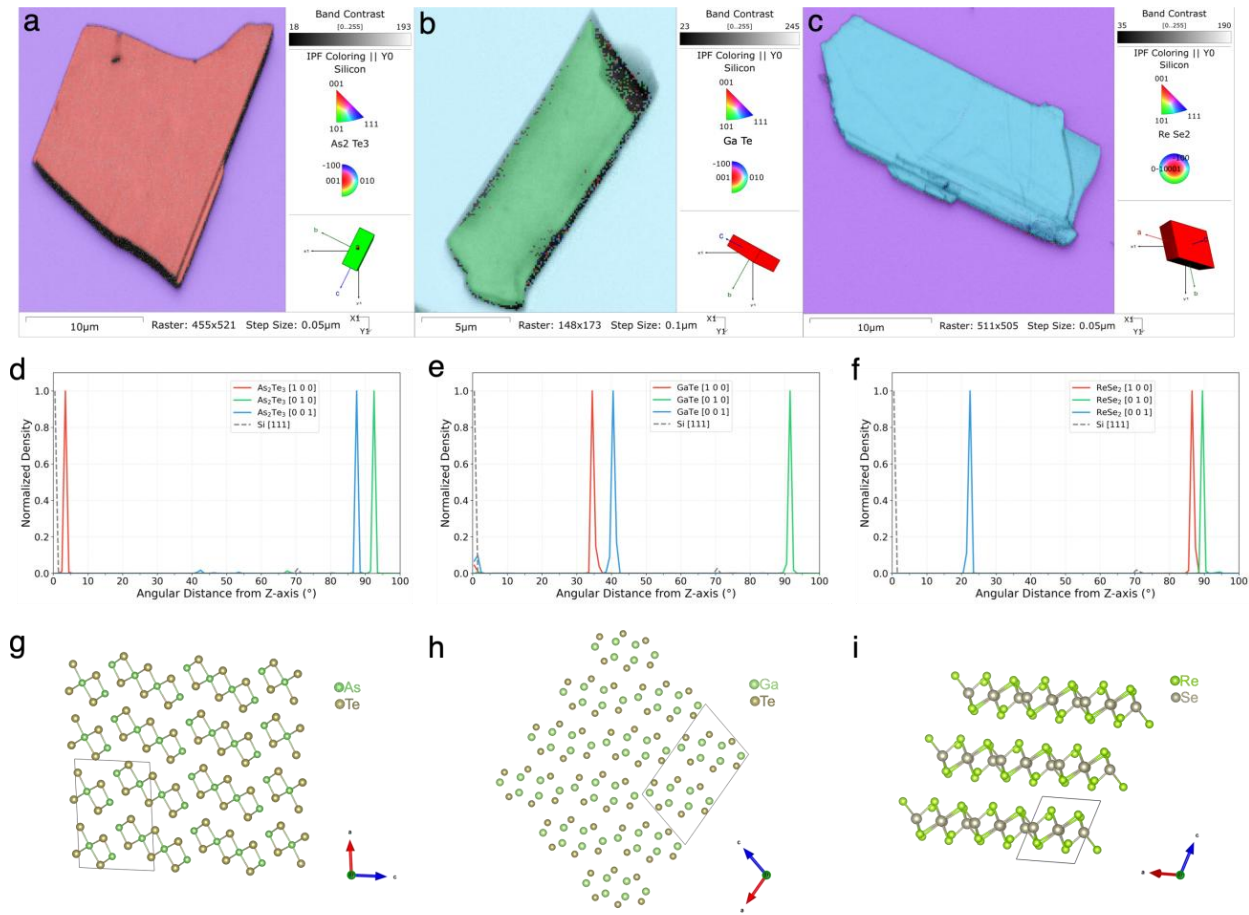

**Figure 3. EBSD of lower symmetry vdW flakes. (a–c)** Inverse Pole Figure (IPF-Y) maps overlaid with band contrast for (a) α-As$_2$Te$_3$, (b) GaTe, and (c) ReSe$_2$. The color keys indicate the crystallographic orientation relative to the Y-axis of sample reference frame. The crystallographic parameters used are: α-As$_2$Te$_3$ (a=14.34 Å, b=4.01 Å, c=9.87 Å; α=γ=90°, β=95°); GaTe (a=17.4 Å, b=4.08 Å, c=10.46 Å; α=γ=90°, β=104.5°); and ReSe$_2$ (a=6.6 Å, b=6.71 Å, c=6.72 Å; α=91.84°, β=104.9°, γ=118.91°). The insets show the unit cell orientations at representative points on the flakes. **(d–f)** Corresponding normalized density plots showing the angular distribution of the primary crystal axes ([100], [010], and [001]) relative to the surface normal (Z-axis). The dashed grey lines represent the orientation of the Si [111]

substrate aligned with the Z-axis. The plots reveal distinct out-of-plane and in-plane unit cell orientations for each material. **(g-i)** Approximate side views of the crystal structure of the measured flakes looking down on the [010] axis constructed by aligning their [001] axis using the values shown in (d-f).

**Table 1.** KAM and GROD statistics in lower symmetry vdW crystal samples.

| Sample | KAM Mean (°) | KAM Median (°) | GROD Mean (°) | GROD Median (°) |
|---|---|---|---|---|
| $As_2Te_3$ | 0.13 | 0.11 | 0.16 | 0.13 |
| GaTe | 0.25 | 0.20 | 0.35 | 0.23 |
| $ReSe_2$ | 0.10 | 0.08 | 0.18 | 0.12 |

As a proof-of-concept application of this high-precision crystal orientation characterization, we fabricated a twisted α-MoO₃ bilayer designed to exhibit directional phonon polariton (PhP) propagation. Polaritons are hybrid quasiparticles resulting from the strong coupling between photons and dipole-active matter excitations. They provide a pathway to confine light to deep sub-wavelength scales. Of particular interest are hyperbolic polaritons, which arise in anisotropic media where the dielectric permittivity possesses opposite signs along different principal axes, enabling the propagation of modes with directionality and extremely high momenta. Among vdW materials, α-MoO₃ has emerged as a natural hyperbolic material, hosting PhPs that exhibit in-plane anisotropy. While the propagation of PhPs in a single crystal is dictated by its intrinsic crystallographic axes and the excitation frequency, twisted bilayer systems introduce a new degree of freedom for dispersion engineering. By stacking two α-MoO₃ slabs with a relative rotational misalignment, the hybridization of modes modifies the topology of the polaritonic isofrequency contours, allowing for the tuning of dispersion between open hyperbolic and closed elliptic regimes. Critically, at specific twist angles, this topological transition enables canalization, a regime where the isofrequency curve flattens, suppressing diffraction and forcing energy to flow in a highly collimated beam. Achieving this regime at a specific frequency requires a specific relative twist angle between the two layers, often within a tolerance of 5° [24].

Using EBSD, we pre-characterized two individual α-MoO₃ flakes, identifying their crystallographic axes with high precision with respect to the orientation of the Si substrate which were seen to align with their facet orientation. Based on this data, we deterministically stacked the flakes using dry transfer to achieve a targeted twist angle of 70° on Si substrate. Gallium ions are used in a FEI Helios NanoLab G3 CX focused-ion beam (FIB) to mill ~500 nm diameter through hole that is used to launch polaritons in the bilayer region. Figure 4(a) shows an electron image of the sample with a 270 nm top flake stacked on top of a 100 nm bottom flake along with the FIB hole. We then performed EBSD on the heterostructure to confirm the twist angle between the two α-MoO₃ flakes (Figure 4(b)). By mapping the orientation of the exposed bottom layer and the top layer within the same scan, we calculated the physical twist angle to be 71.74° directly from the relative rotation of their Euler angles (Figure 4(c)). This is crucial in some applications as vdW stacks are known to show lattice relaxation after fabrication [50, 51]. The deviation from the target angle in our sample can be largely attributed to limitations of the transfer setup.

Subdiffractional near-field optical imaging of the fabricated structure is then carried out using a commercial Neaspec system illuminated using a continuous wave MIRcat quantum cascade laser (QCL). The third harmonic of optical signal from the scattering-type Scanning Near-field Optical Microscopy (s-SNOM) (Figure 4(d-f)) contains primarily near field signals from the PhPs when using a pseudo-heterodyne configuration [52]. Images include a background correction described in the Supporting Information (Figure S4), which minimizes the contribution from the edge launched PhPs of the top flake. This helps us isolate the hybridized PhPs launched from the FIB hole, revealing the elliptical, hyperbolic and canalized PhP propagation with changing frequency. We observe characteristic directional propagation pattern associated with canalization at 885 $cm^{-1}$ as confirmed by the 2D Fast Fourier Transform (FFT) of the field (Figure 4(g)). This phenomenon is reported in literature with frequencies around 900 $cm^{-1}$ for different thicknesses and twist angles on $SiO_2$/Si substrates [21, 22, 24]. The real part of the electric field obtained from Finite Element Method (FEM) simulations (Figure 4(h)) performed using the twist angle determined by EBSD on the stack agree with the experimental observation. This post-fabrication validation highlights the distinctive capability of EBSD to probe the crystallographic information of assembled vdW devices, providing a true structural verification that complements functional optical characterization. We note that electron exposure during EBSD has not hindered the polaritonic properties of the material. Thus, EBSD provides a direct and quantitative pathway to optimize and fabricate vdW heterostructures with twist-dependent optical functionalities and perform post-fabrication verification of the twist angle between flakes, positioning it as a valuable tool for the development of twist-optics.

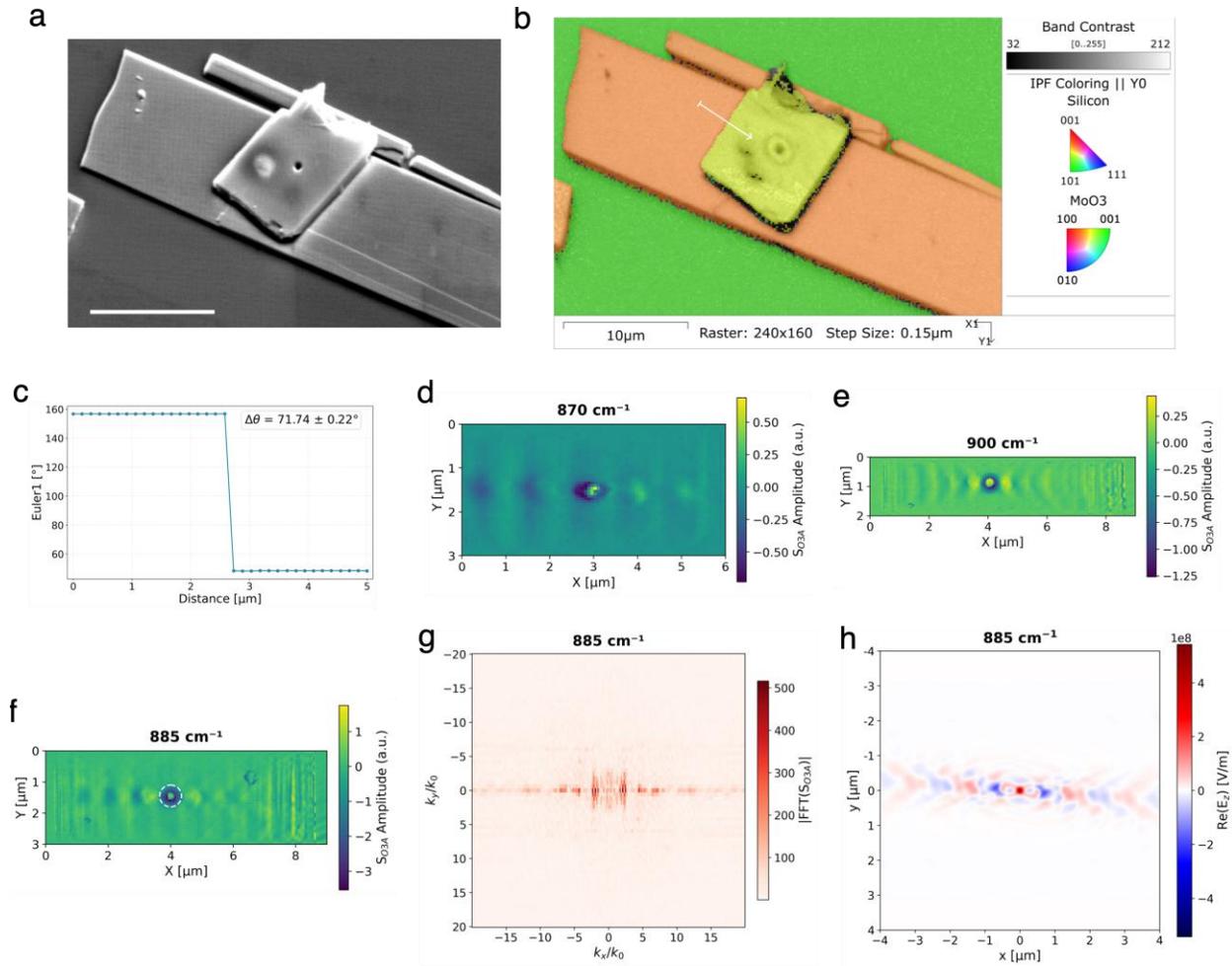

**Figure 4. PhPs in twisted α-MoO₃ with EBSD characterization.** (a) SEM image of the twisted α-MoO₃ stack on Si. The scale bar is 10 µm. (b) Inverse Pole Figure (IPF-Y) orientation map overlaid with band contrast for the twisted stack, showing the orientations of the two flakes. The white line represents the line measurement shown in (c). (c) Line measurement of the Euler angle corresponding to the in-plane orientation of [001] crystal axis, showing a twist angle of 71.74°. (d-f) Near field amplitudes measured with s-SNOM showing elliptical, hyperbolic and canalized Phonon Polaritons (PhPs) at 870 cm$^{-1}$, 900 cm$^{-1}$ and 885 cm$^{-1}$ respectively. The Y- axis is approximately aligned with [001] axis of the top flake. (g) 2D FFT of the field at 885 cm$^{-1}$ shown in (f) excluding the region within the dotted circle. k₀ is the momentum of infrared photon, $k_0 = 2\pi/\lambda_0$ (h) FEM field simulation showing canalized PhPs at 885 cm$^{-1}$ in the near field with [001] axis of the top flake set to vertical and that of the bottom flake set to 71.74° clockwise with respect to the top.

In conclusion, we establish EBSD as a robust, accurate, and broadly applicable method for determining the crystallographic orientation of vdW materials. The technique shows excellent accuracy as demonstrated through agreement between EBSD-determined orientations and facet alignment in orthorhombic α-MoO₃ flakes as a benchmark, while achieving an angular precision of few tenths of a degree as determined through KAM analysis across multiple flakes. Furthermore, EBSD provides reliable and precision orientation mapping in vdW materials with lower symmetries as shown in our work for monoclinic and triclinic materials where facet-based alignment is often unreliable or impossible. We also observe that the vdW flakes studied in this work show minimal lattice deformation from GROD analysis. As a practical demonstration, we then employ EBSD-guided deterministic stacking to fabricate a twisted α-MoO₃ heterostructure that supports canalized phonon polariton propagation, with post-fabrication EBSD providing direct verification of the final twist angle. These results establish EBSD as an effective and practical structural characterization tool for high precision twistronics and twist-optics in vdW stacks.

## Supporting Information

See the supporting information for SEM images of analyzed α-MoO₃ flakes, EBSD measurement parameters, details of facet orientation determination, additional KAM and GROD data, s-SNOM background correction method, AFM thickness characterization, and FEM simulation details.

## Acknowledgements

RB, TGF, JB and JDC acknowledge support from the Multi-University Research Initiative (MURI) on Twist-Optics, sponsored by the Office of Naval Research (ONR) under Grant No. N00014-23-1-2567. MM and NB thank the Natural Sciences and Engineering Research Council of Canada (NSERC) for financially supporting this work under the Alliance International Catalyst Quantum grants program (ALLRP 580935 – 22). The authors also acknowledge support for carrying out the Electron Backscatter Diffraction (EBSD) at the Canadian Centre for Electron Microscopy (CCEM), a national facility supported by McMaster University, the Ontario Research Fund (ORF), and the Canada Foundation for Innovation (CFI). Some of the work was carried out at the University of Iowa Materials Analysis, Testing, and Fabrication (MATFab) facility.

## Conflict of Interest

The authors have no conflicts to disclose.